\documentclass[aps,prl,groupedaddress]{revtex4}
\usepackage{graphicx}
\usepackage{here}
\usepackage{algorithm}
\usepackage{braket}
\usepackage[noend]{algpseudocode}
\usepackage{amsmath}
\usepackage{amssymb}
\begin{document}

\preprint{APS/123-QED}

\title{Successive generation of nontrivial Riemann zeros from a Wu-Sprung type potential}

\author{Peter Jaksch}
\email{peter.jaksch@gmail.com}

\date{\today}

\begin{abstract}
A series of numerical experiments are performed, where a symmetric potential is generated for the 1D time-independent Schrödinger equation, with an eigenspectrum that matches the imaginary part of the first  nontrivial zeros of the Riemann Zeta Function. 
The potential is generated as a series of correction functions, where the starting point 
is a potential that matches the smooth Riemann -- von Mangoldt approximation.
It is found that the correction functions display a clear pattern that can be explained in simple terms,
almost entirely dependent on the approximation error in the Riemann -- von Mangoldt formula. This also provides an explanation for the fractal pattern in the potential that was observed in \cite{wu-sprung}.
\end{abstract}

\maketitle

\section{introduction}

According to the Hilbert - Pólya conjecture; if $H$ is a self-adjoint operator and the eigenvalues of $1/2 + iH$ correspond to the 
nontrivial zeros of the Riemann zeta function, then the Riemann hypothesis \cite{riemann} follows. Important efforts to find such an operator have
been done by e.g. \cite{keating} -- the so-called Berry-Keating conjecture, and later by \cite{bender} (see \cite{quanta} for an excellent introduction). In the latter case, the proposed operator is of the form

\begin{equation}
\hat{H} = \frac{1}{1-e^{-i\hat{p}}} (\hat{x}\hat{p} + \hat{p}\hat{x}) (1-e^{-i\hat{p}}),
\end{equation}
where $\hat{p}$ and $\hat{x}$ are the quantum mechanical momentum and position operators, respectively.
Although the eigenvalues of the operator match the nontrivial eigenvalues, 
it has not yet been possible to rigorously prove that the operator is self-adjoint. 

An alternative approach is to instead work with a self-adjoint operator on a specified domain and try to modify it's eigenvalues to match the Riemann zeros. This approach has been 
taken in \cite{wu-sprung}. A similar approach is taken in the present paper. The operator is the Hamiltonian of the 1D time-independent Schrödinger on $\mathbb{R}$, which, in suitable units, can be 
written as an eigenvalue problem
\begin{equation}
 \left( -\frac{\partial^2}{\partial x^2} + V(x) \right) \Psi(x) = E \Psi(x) \label{1d}
\end{equation}
The potential $V(x)$ is assumed to be mirror symmetric around $x=0$. If the kinetic term is denoted $T$, equation (\ref{1d}) can be written more compactly as $(T + V) \Psi = E \Psi$.

Instead of directly trying to match the Riemann zeros with the eigenvalue spectrum, it is easier to first match a smooth approximation to the zeros.
Various such approximations have been derived (see e.g. \cite{tim}). In the present paper the earliest such approximation is used, referred to as the 
Riemann -- von Mangoldt formula \cite{riemann}. It states that the number $N(T)$ of zeros of the zeta function with imaginary part greater than 0 and less than $T$
satisfies

\begin{equation}
N(T) = \frac{T}{2 \pi} \log  \frac{T}{2 \pi} -  \frac{T}{2 \pi} + O(\log T) \label{RM}
\end{equation}

\begin{figure}[H]
\centering
\includegraphics[width=12cm]{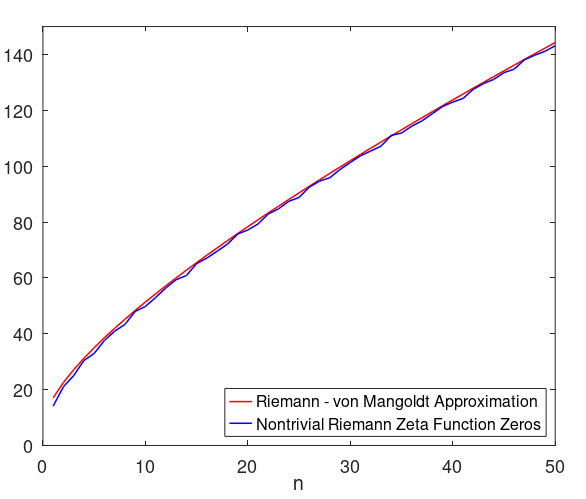}
\caption{The imaginary part of the first 50 nontrivial Riemann zeros and the Riemann - von Mangoldt smooth approximation.}
\label{smooth}
\end{figure}

This function has to be inverted if, instead, the approximate location of each zero is of interest. 
\begin{equation}
Z_n = \frac{2n}{\pi W(n/e)}, \label{lambert}
\end{equation}
where $W$ is the Lambert W-function. FIG. \ref{smooth} shows the approximation for the first 50 zeros. For brevity, the imaginary part of the nontrivial Riemann zeros
in the upper half-plane will henceforth be referred to as simply; Riemann zeros. For later use it should be noted that the only zero that is larger than the approximation in FIG. \ref{smooth} is $Z_{34}$.

\section{a potential for the smooth spectrum}

\begin{figure}[H]
\centering
\includegraphics[width=15cm]{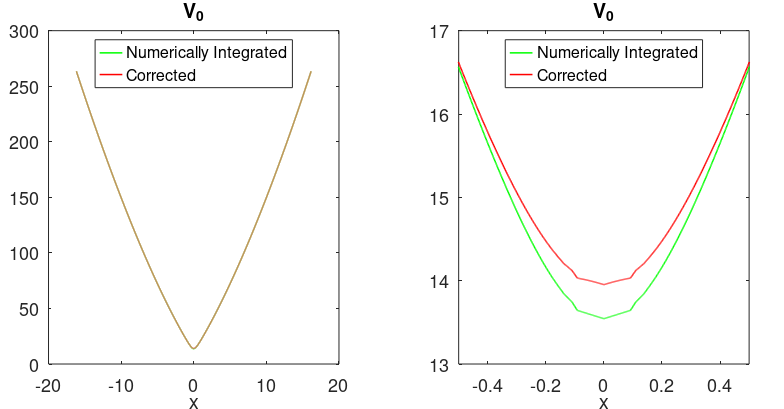}
\caption{Smooth potential with eigenvalues matching the approximate location of the nontrivial Riemann zeros, calculated from the Riemann - von Mangoldt formula. 
The figure on the right  provides a zoomed-in view.}
\label{potential0}
\end{figure}

It is possible to derive a potential $V(x)$ in (\ref{1d}) whose eigenvalue spectrum matches (\ref{lambert}). In \cite{wu-sprung} this is done analytically for a slightly different smooth 
approximation, by assuming a semiclassical approximation, leading to an Abel integral equation. The present paper takes a similar approach but instead solves the problem numerically.
Using the semiclassical  WKB approximation (see \cite{wkb}) the eigenvalue problem (\ref{1d}) gives the following relation

\begin{equation}
\int_{-x_0}^{x_0} \sqrt{E - V(x)} dx = \left(n + \frac{1}{2} \right) \pi, \label{wkb-eigenvalues}
\end{equation}
where $E$ is the energy level, $x_0$ is the classical turning point where $V(x_0) = E$, and $n=0,1,2,\dots$ is a quantum number. Combining (\ref{wkb-eigenvalues}) with (\ref{RM}) gives
\begin{equation}
\int_{-x_0}^{x_0} \sqrt{E - V(x)} dx = \frac{E}{2} \left( \log \left( \frac{E}{2 \pi} \right) - 1 \right) + \frac{\pi}{2}. \label{energy}
\end{equation}
An increase $\Delta E$ in energy gives a corresponding increase in the turning point $\Delta x_0$, such that $V(x_0 + \Delta x_0) = E + \Delta E$. 
From this relation, the shape of the potential can be calculated numerically if 
the starting value $V(0)$ is known. By letting $f(E)$ denote the right hand side of (\ref{energy}), and utilizing the assumption of mirror symmetry, an equation for the differences can 
be formulated
\begin{equation}
2 \int_{0}^{x_0 + \Delta x_0} \sqrt{E + \Delta E - V(x)} dx \approx f (E + \Delta E).
\end{equation}
Due to the smoothness of the problem it can be assumed that $V(x)$ can be linearized on the interval $[x_0, x_0+\Delta x_0]$. Then $\Delta x_0$ can be calculated as 
\begin{equation}
\Delta x_0 \approx \frac{3}{4 \sqrt{\Delta E}} \left( f (E + \Delta E) - 2 \int_{0}^{x_0} \sqrt{E + \Delta E - V(x)} dx  \right). \label{x-diff}
\end{equation}
If the starting point $V(0)$ is known, the above procedure can be repeated, with a chosen step-size $\Delta E$, in order to generate the entire potential $V(x)$ for an arbitrary value of $x$.
From (\ref{energy}) it is clear that the left hand side of the equation is 0 if $E = V(0)$. The right hand side of (\ref{energy}) can be solved numerically for finding the value of $E$ such that $f(E)=0$.
With a suitable solution method this value can be calculated to be $E \approx 13.544$, which is also the value of $V(0)$. FIG. \ref{potential0} (green line) shows the shape of the potential, 
generated by this method. It is relatively similar to a harmonic oscillator. This is not too surprising given that the spacing between the eigenvalues is relatively constant -- a hallmark 
of the quantum harmonic oscillator.

It is possible to improve the accuracy of the matching of the eigenvalue spectrum by utilizing a clever variational method, devised in \cite{wu-sprung}. The method is repeated here for clarity.
Let $e_n$ be the actual eigenvalues of (\ref{1d}) and $E_n$ the target eigenvalues. The least squares function to minimize is 
\begin{equation}
F = \sum_n (e_n - E_n)^2. \label{LS}
\end{equation}
The functional derivative of $e_n$ with respect to $V(x)$ is 
\begin{equation}
\frac{\delta e_n}{\delta V(x)} = \frac{\delta}{\delta V(x)} \bra{\Psi_n} T + V \ket{\Psi_n} = \Psi_n(x)^2,
\end{equation}
where $\Psi_n(x)$ is the normalized eigenfunction corresponding to eigenvalue $n$. For the least squares expression, this gives:
\begin{equation}
\frac{\delta F}{\delta V(x)} = 2 \sum_n (e_n - E_n) \Psi_n(x)^2. \label{var}
\end{equation}
Normally, the problem is discretized on a set of grid points, in which case (\ref{var}) becomes a gradient vector. A finite difference scheme is used to discretize the operator $T$:
\begin{equation}
T = -\frac{1}{\Delta x^2} 
\begin{pmatrix}
-2 & 1 & &  \cdots \\
\hspace{3mm} 1 & \hspace{-3mm}  -2  & \hspace{-8mm} 1 & \cdots \\
\vdots & \hspace{8mm}   \ddots & & \vdots \\
\cdots & \hspace{8mm} 1 & -2 & \hspace{3mm} 1 \\
\cdots &  & \hspace{3mm} 1 &  -2
\label{finite-diff}
\end{pmatrix}
\end{equation}
Note that Dirichlet boundary conditions at the endpoints have been implemented implicitly here. This will not cause any problems since, for the lower eigenfunctions that are of interest here, 
decay to 0 is rapid near the endpoints. In this case a regular grid with 20 000 points has been used. By also discretizing the potential on the same grid, the eigenproblem (\ref{1d})
can be turned into a standard eigenproblem for (sparse) matrices, for which there are efficient software implementations.
The actual minimization of (\ref{LS}) can be performed by 
utilizing the gradient vector in an optimization algorithm. In this case, a conjugate gradient minimizer, together with a line search method, has been implemented in a MATLAB 
script. The adjustment of the potential, calculated by (\ref{x-diff}), can be seen in FIG. \ref{potential0} (red line).

\section{successive matching of Riemann zeros}

\begin{figure}[H]
\centering
\includegraphics[width=14cm]{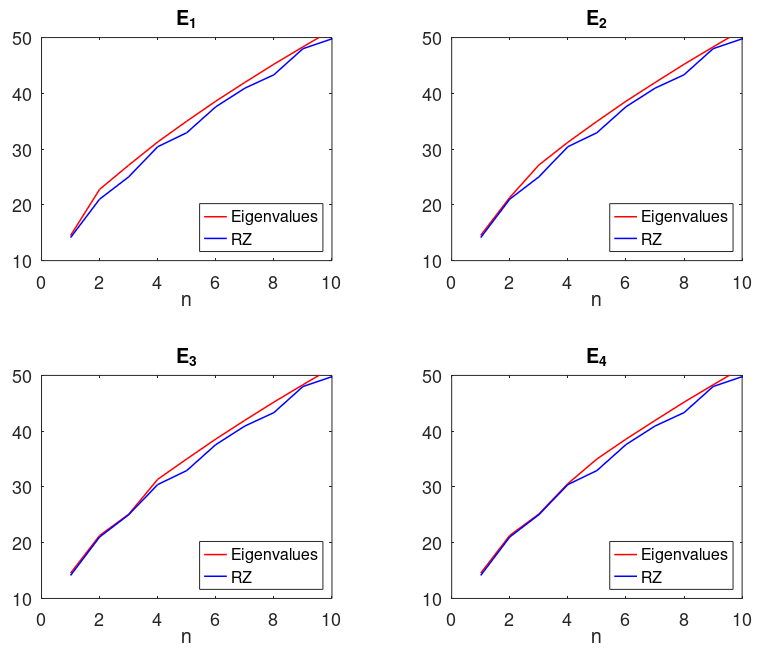}
\caption{Eigenvalue spectrum of the potentials in FIG. \ref{potentials_1_4}.}
\label{spectrum_1_4}
\end{figure}

\begin{figure}[H]
\centering
\includegraphics[width=12cm]{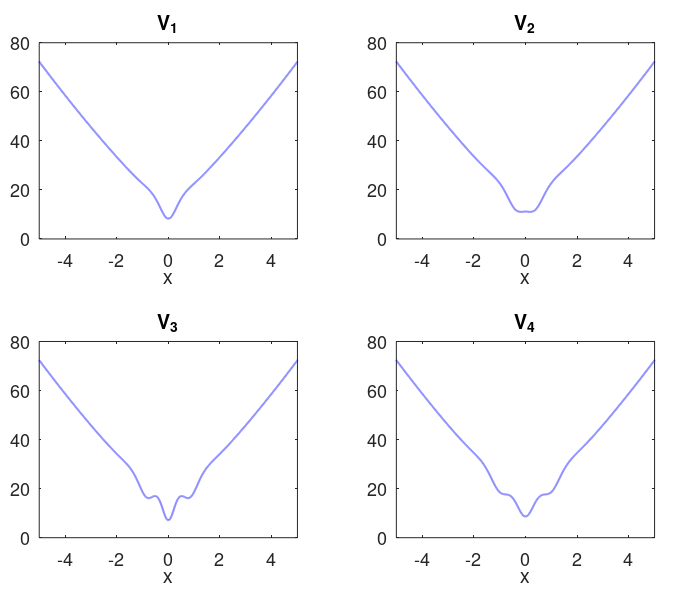}
\caption{Successive corrected potentials for matching the first four nontrivial Riemann zeros.}
\label{potentials_1_4}
\end{figure}

The tools and methods developed so far can be used to match the actual Riemann zeros; not just the smooth approximation. In the present paper this has been performed
with a step-wise approach where one extra Riemann zero is matched in every new iteration, hoping that some identifiable pattern will emerge. 
This differs from the approach taken in \cite{wu-sprung} where a few sets with
a large number of eigenvalues were matched at once. The results for the first four Riemann zeros is illustrated in FIG. \ref{spectrum_1_4} -- FIG. \ref{potentials_1_4}.
The terminology is chosen such that 
\begin{equation}
V_n(x) = V_0(x) + \sum_{i=1}^{i=n} C_i(x), \label{v-reconstruct}
\end{equation}
where $V_0$ is the corrected potential, derived in the previous section; $V_n$ is the potential with an eigenspectrum that matches the first $n$ Riemann zeros,
and the Riemann -- von Mangoldt approximation for higher eigenvalues. The potentials $V_1 -V_{50}$ are calculated numerically, using the approach in the previous section, with $C_n$ being postprocessed as $C_n=V_{n+1}-V_n.$

\begin{figure}[H]
\centering
\includegraphics[width=12cm]{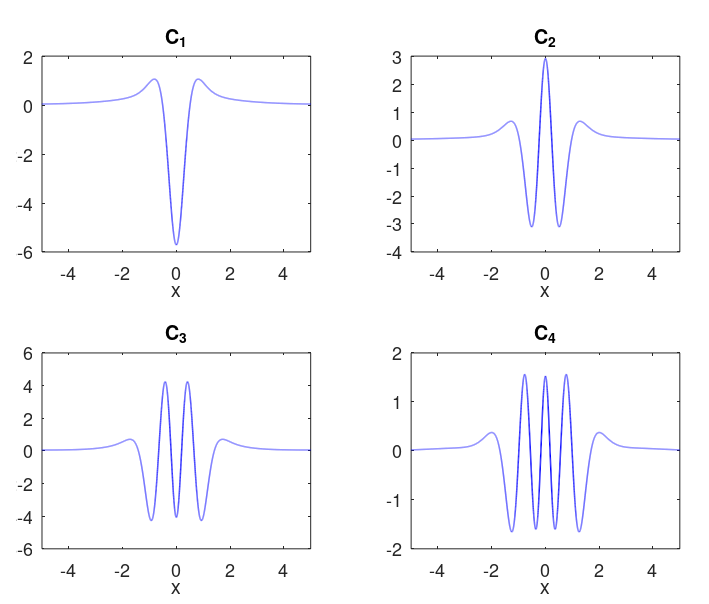}
\caption{Successive corrections to the smooth potential $V_0$ in FIG. \ref{potential0} for matching the first four nontrivial Riemann zeros.}
\label{corrections_1_4}
\end{figure}

\begin{figure}[H]
\centering
\includegraphics[width=8cm]{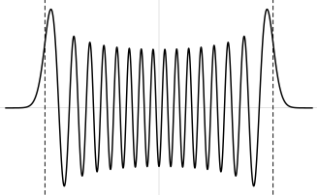}
\caption{A symmetric eigenfunction for the quantum harmonic oscillator.}
\label{harmonic}
\end{figure}

It can be seen in FIG. \ref{corrections_1_4} that the corrections follow a surprisingly regular pattern.
It seems to consist of a sine wave, symmetric around both axes, with constant amplitude, followed by exponentially decaying tails on either side. $C_1(x)$ consists of two tails, joined at $x=0$; $C_2(x)$ 
has a single period of a sine wave between the tails; $C_3(x)$ has two periods, etc. It also seems like the wavelength increases farther away from the origin. Overall,
the correction functions look remarkably similar to the (symmetric) eigenfunctions of the quantum harmonic oscillator (see FIG.  \ref{harmonic}), albeit with a constant amplitude
instead of an increasing amplitude towards the classical turning points (dashed lines).

\begin{figure}[H]
\centering
\includegraphics[width=10cm]{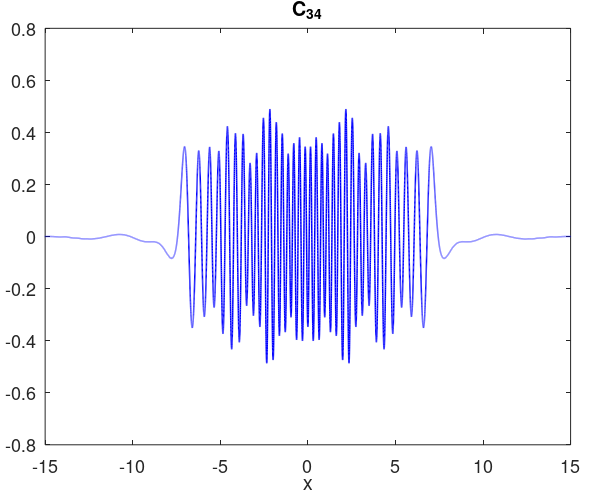}
\caption{The only correction, out of the first 50, with a different sign.}
\label{c34}
\end{figure}

Only one of the first 50 correction stands out by the fact that it seems to have a different sign. This is $C_{34}$ (see FIG. \ref{c34}). It was previously noted that this number was also 
the only Riemann zero where the Riemann - von Mangoldt approximation was below the actual value. This leads to the suspicion that the amplitudes of the oscillating part
of the correction functions is related to the smooth approximation error of a given Riemann zero. 

 $C_{34}$ also displays quite large variations in amplitude. This could be a real effect but it may also be due to numeric noise, or the fact that inverse eigenproblems are notorious for the 
fact that they often lack a unique solution. In order to investigate if the variation in amplitude increases for higher corrections, $C_{50}$ is shown below. Again,
the amplitude is almost constant. Manual inspection shows that most correction functions seem to have only small variations in amplitude. In fact, this was the main reason
for using the Riemann -- von Mangoldt approximation, which is generally less accurate than more modern alternatives, such as the one used in \cite{wu-sprung}. 

\begin{figure}[H]
\centering
\includegraphics[width=10cm]{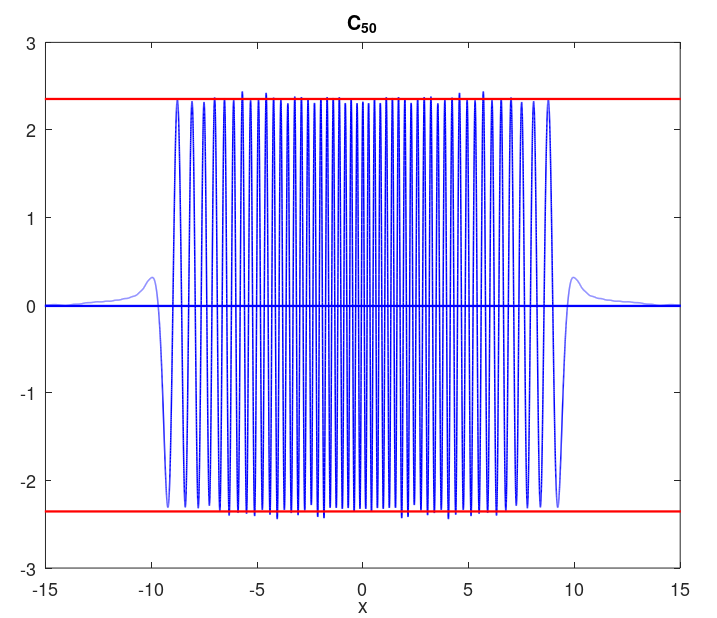}
\caption{Correction number 50. Note the almost constant amplitude.}
\label{c50}
\end{figure}

\begin{figure}[H]
\centering
\includegraphics[width=13cm]{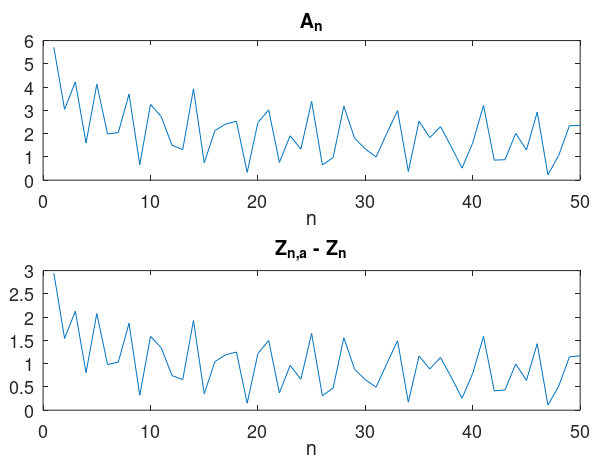}
\caption{The amplitude of corrections $C_1$ to $C_{50}$ shows a strong correlation with the Riemann - van Mangoldt approximation error.}
\label{amplitude_correlation}
\end{figure}

 \newpage

\begin{figure}[H]
\centering
\includegraphics[width=10cm]{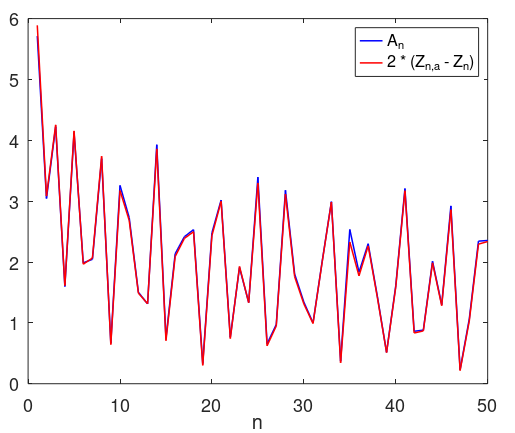}
\caption{The correction amplitudes can be accurately recreated by doubling the Riemann - von Mangoldt approximation error.}
\label{amplitude_reconstruction}
\end{figure}

Based on the observation in the previous section, that the amplitude of the oscillating part of the correction functions seems correlated with the smooth approximation error,
these two properties are plotted together in FIG. \ref{amplitude_correlation}. The amplitude is calculated numerically as the average peak value of the oscillating part of the correction.
Indeed there is an obvious strong correlation. By doubling the approximation error, the numerically calculated amplitudes are recovered almost perfectly (see FIG. \ref{amplitude_reconstruction}).  In summary
\begin{equation}
A_n = 2 (Z_{n, a} - Z_n),
\end{equation}
where, $A_n$ is the amplitude of the oscillating part of $C_n$, $Z_n$ is Riemann zero $n$, and $Z_{n,a}$ is the Riemann -- von Mangoldt approximation to Riemann zero $n$. This seems to
hold, even for functions with a relatively large variation in amplitude, such as $C_{34}$. It was also found to hold if the smooth approximation from \cite{wu-sprung} was used.
The fact that the amplitude can be explained so well is rather unexpected.

\begin{figure}[H]
\centering
\includegraphics[width=15cm]{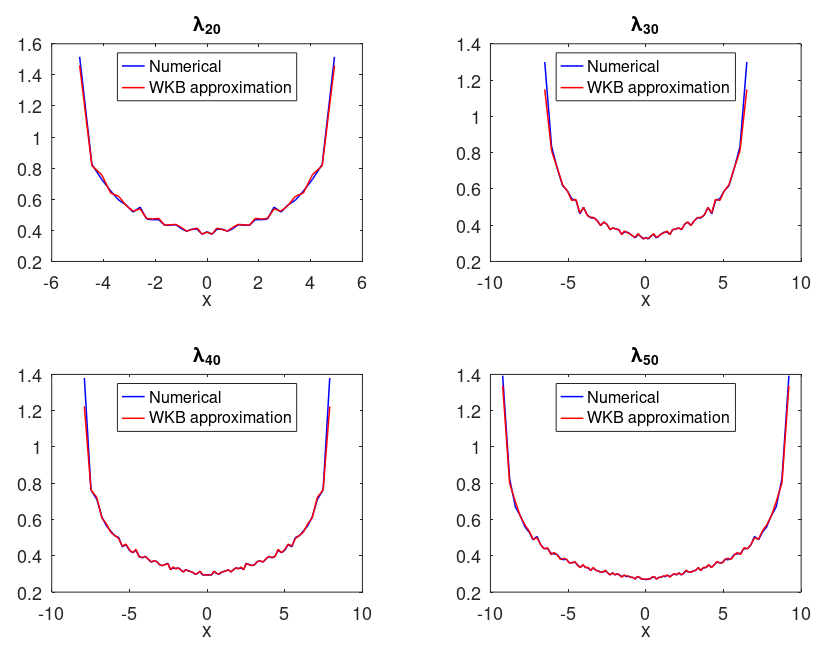}
\caption{The local wavelengths of corrections $C_{20}, C_{30}, C_{40}, C_{50}$, calculated numerically and with the WKB approximation.}
\label{wavelengths}
\end{figure}

The similarities with the quantum harmonic oscillator provides some hint on how to explain the local frequency of the periodic part of the correction. The semiclassical 
WKB approximation \cite{wkb} is known to provide exact solutions for the quantum harmonic oscillator, on which the smooth potential in FIG. \ref{potential0} is reminiscent. 
The local angular frequency from the WKB approximation is proportional to the semiclassical momentum $p=\sqrt{E - V(x)}$, for a particle with a given energy $E$.
The local wavelength for the correction functions is calculated by first finding the local max and min points, and then locating the two closest points where the function crosses 0. 
The local wavelength is twice the distance between the zero crossings. This procedure works best for the functions with higher wave numbers. FIG. \ref{wavelengths}
shows numerical results for $C_{20}$, $C_{30}$, $C_{40}$, and $C_{50}$. In the same figure it can be seen that excellent agreement with the numerical method
can be obtained by doubling the WKB frequency. The local wavelength is then defined as
\begin{equation}
\lambda_n(x) = \frac{\pi}{\sqrt{Z_n - V(x)}}, \label{lambda-wkb}
\end{equation}
where $Z_n$ is the corresponding Riemann zero. 
With the results, obtained to far, it is now possible to reconstruct the oscillating part of the correction functions:
\begin{equation}
C^+_n(x) = 2 (-1)^n (Z_{n,a}-Z_n) \cos \left( \int_0^x \sqrt{Z_n - V_n} dx \right). \label{reconstruct}
\end{equation}
To generate $C^-_n(x)$ for negative values of $x$, the mirror symmetry condition can be used.

\begin{figure}[H]
\centering
\includegraphics[width=10cm]{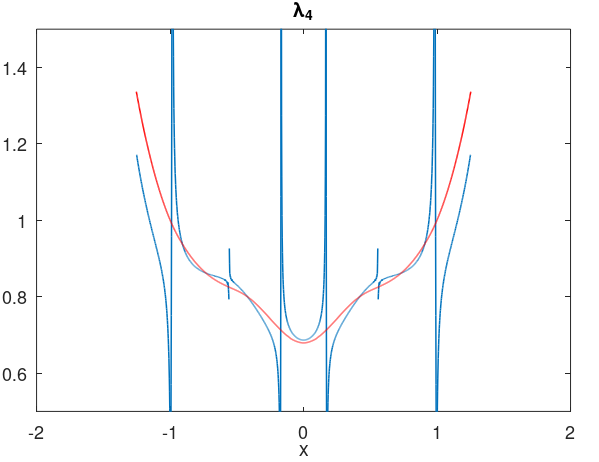}
\caption{The local wavelengths of corrections $C_{4}$, calculated numerically using the second derivative operator (blue line) , and with the WKB approximation (red line).}
\label{wavelengths_4}
\end{figure}

For lower wavenumbers, the fact that $\frac{d^2}{dx^2} \sin{(\omega x)} = -\omega^2  \sin{(\omega x)}$ can be used to 
calculate the wavelength:
\begin{equation}
\lambda_n(x) = \frac{2 \pi}{\sqrt{|C_n''(x) / C(x)|}} \label{lambda}
\end{equation}
This method is not numerically stable around the points where $C(x)=0$. In FIG. \ref{wavelengths_4} $\lambda_n(x)$, calculated using (\ref{lambda}), is plotted against
the WKB approximation (\ref{lambda-wkb}). Also in this case the agreement is good, except in the vicinity of the singularities.

\begin{figure}[H]
\centering
\includegraphics[width=10cm]{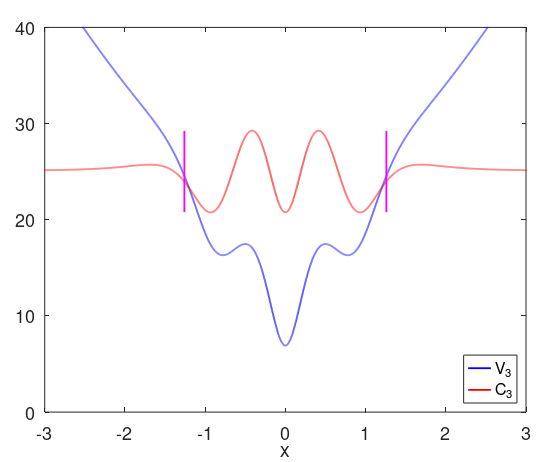}
\caption{Potential $V_3$ plotted together with correction $C_3$ and the semiclassical turning points .}
\label{turning}
\end{figure}

In order to construct the functions $C_n(x)$ on the entire domain of $x$-values, a reconstruction of the tail section is needed. From the graphs of various $C$-functions
shown so far, it seems like the tail section can be described as a skewed exponential function. At first, this seems, again, similar to what happens in the classically forbidden
region in the WKB approximation (see FIG. \ref{harmonic}). However, a plot with the classical turning points marked (FIG. \ref{turning}) shows that they are in a different location, 
compared to FIG. \ref{harmonic}. After passing the classical turning points, the function should decay exponentially to 0. This is not the case in FIG. \ref{turning}. Without a significant modification of the potential it does not seem possible to use the WKB approximation for the tail section.
This problem has caused considerable difficulty. Fortunately, it may be possible to work around the problem without finding a simple explanation for how to generate the tails.

\begin{figure}[H]
\centering
\includegraphics[width=12cm]{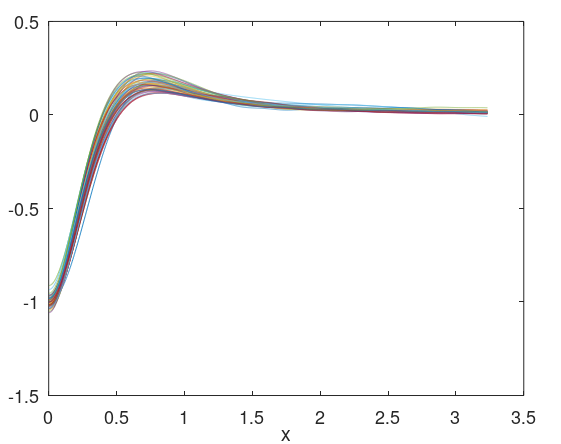}
\caption{Plume of all tails, normalized with amplitude, of $C_1 - C_{50}$ and shifted to $x=0$ .}
\label{tails}
\end{figure}

\begin{figure}[H]
\centering
\includegraphics[width=10cm]{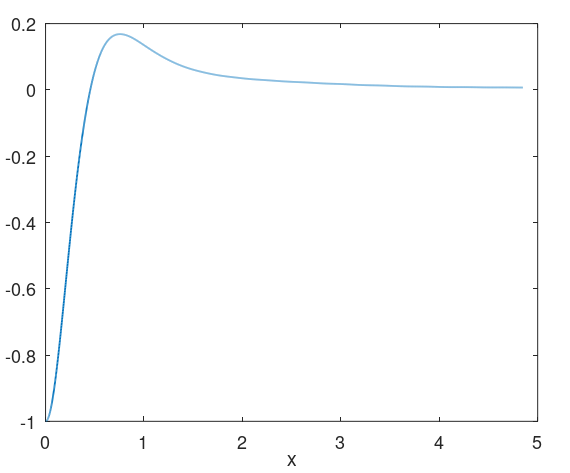}
\caption{Average of all tails in FIG. \ref{tails} and rescaled to start at -1.}
\label{tail}
\end{figure}

FIG. \ref{tails} shows the tails of all 50 functions, normalized with the amplitude of the oscillating part, and shifted to $x=0$. 
It seems like the tails are rather similar. There is some noticeable spread, but by plotting the tails one-by-one, no clear pattern can be discerned. It seems like
the spread is caused mainly by random noise. From this observation, the tail is defined as the average of the 50 functions in FIG. \ref{tails}, and rescaled slightly in order to start at -1
(see FIG. \ref{tail}). It is interesting to note that the average value of the tail function is very close to 0.

\newpage

\begin{figure}[H]
\centering
\includegraphics[width=15cm]{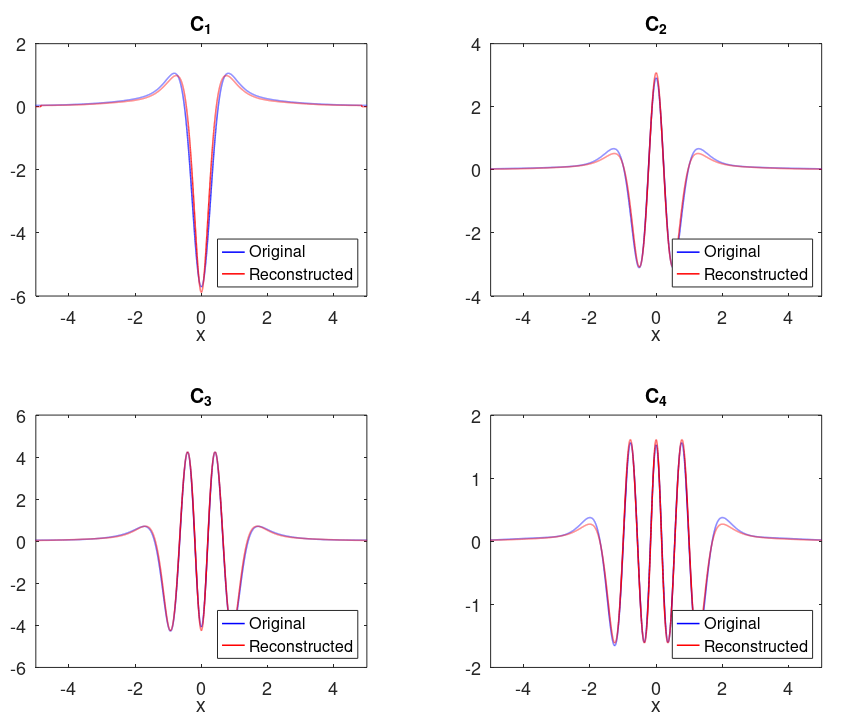}
\caption{Reconstruction of the successive corrections to the smooth potential in FIG. \ref{potential0} for matching the first four nontrivial Riemann zeros.}
\label{reconstruct1-4}
\end{figure}

The function, defined through FIG. \ref{tail}, can now be attached to the endpoints of the oscillating part of the $C$-functions, after rescaling with the amplitude of the oscillating part.
FIG. \ref{reconstruct1-4} shows the result for the first four $C$-functions, where equation (\ref{reconstruct}) has been used for the oscillating part of the functions.
Overall, the agreement seems good.

\begin{figure}[H]
\centering
\includegraphics[width=9cm]{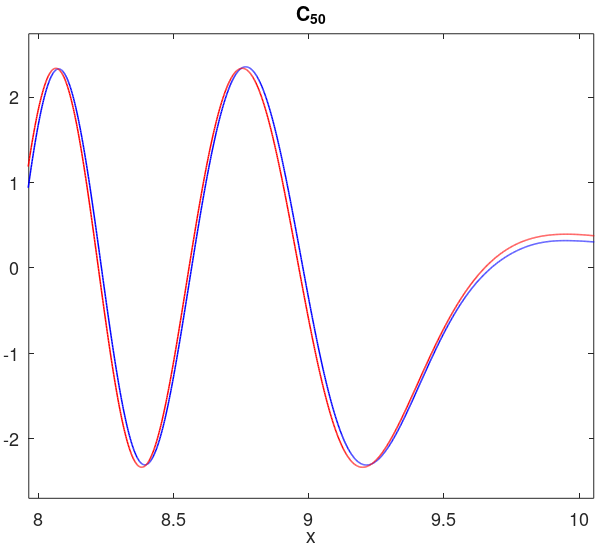}
\caption{The reconstructed function from equation (\ref{reconstruct}) experiences some amount of drift in the frequency.}
\label{drift}
\end{figure}

It is interesting to continue the reconstruction to higher functions. For $C_{50}$ a small but noticeable drift can be seen (see FIG. \ref{drift}).
It is investigated if this shift can be compensated for by modifying equation (\ref{reconstruct}) slightly by adding a small constant shift $s_n$:

\begin{equation}
C^+_n(x) = 2 (-1)^n (Z_{n,a}-Z_n) \cos \left( \int_0^x \sqrt{Z_n - (V_n + s_n)} dx \right). \label{reconstruct_shift}
\end{equation}

\begin{figure}[H]
\centering
\includegraphics[width=12cm]{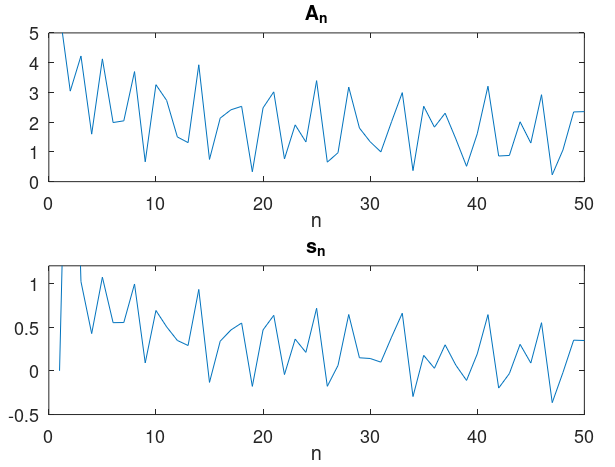}
\caption{Shift for minimizing the drift in the reconstructed functions  $C_1 - C_{50}$ versus amplitude.}
\label{shift}
\end{figure}

The shift can be estimated by minimizing the least squares difference between the oscillating part of the original and reconstructed functions. Since the shift is a constant parameter,
any simple 1D optimization algorithm will work. FIG. \ref{shift} shows the correlation between amplitude and the optimized shift from equation (\ref{reconstruct_shift}). 
It should be noted that this is a correction to a correction of an inverse eigenproblem, so too much cannot be expected in terms of numerical accuracy. Despite this,
it seems to be possible to model the calculated shifts quite accurately, using only the amplitude, as can be seen in FIG. \ref{shift-model}. Manual trial-and-error led to the relation
\begin{equation}
s_n = \frac{A_n-1}{3} \label{shift-model}
\end{equation}

\begin{figure}[H]
\centering
\includegraphics[width=12cm]{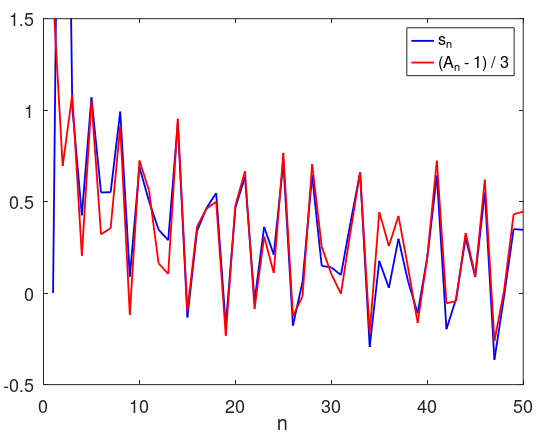}
\caption{The shift $s_n$ for minimizing the drift in the reconstructed functions can be modeled relatively accurately. }
\label{shift-model}
\end{figure}

\section{conclusions}

Summarizing the observations in the previous sections leads to the following properties of $C_n(x)$:
\begin{enumerate}
\item Mirror symmetry around the $y$-axis: $C_n(-x) = C_n(x)$.
\item A periodic section (sine-wave) with constant amplitude $A_n$ with $n-1$ periods.
\item Amplitude $A_n$ depends only on the approximation error in the Riemann -- von Mangoldt formula.
\item The local frequency is proportional to that in the semiclassical WKB approximation, with a small shift that can be modeled in terms of $A_n$.
\item The oscillating part can be described by equations (\ref{reconstruct_shift}) and (\ref{shift-model}).
\item A tail section, with an average value of 0, described by FIG. \ref{tail}, which has to be scaled with the amplitude $A_n$.
\end{enumerate}

\begin{figure}[H]
\centering
\includegraphics[width=15cm]{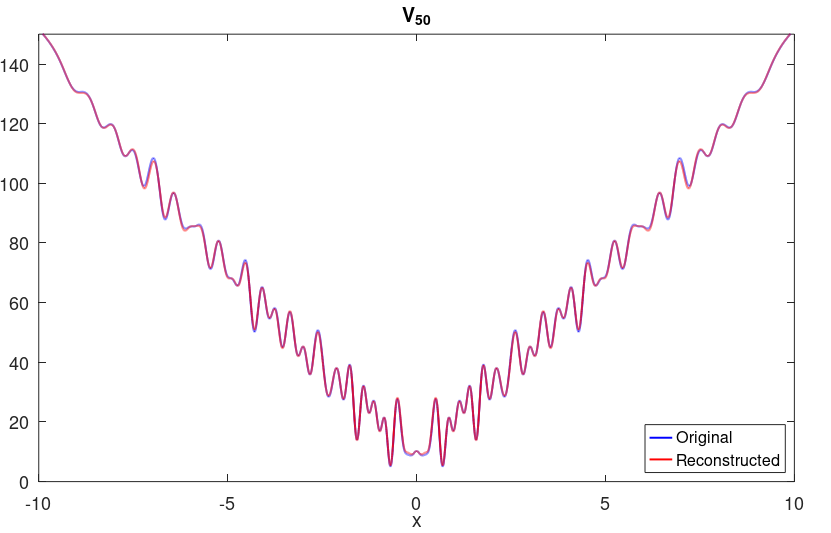}
\caption{Potential with eigenvalues matching the first 50 nontrivial Riemann zeros.}
\label{potential50}
\end{figure}

\begin{figure}[H]
\centering
\includegraphics[width=12cm]{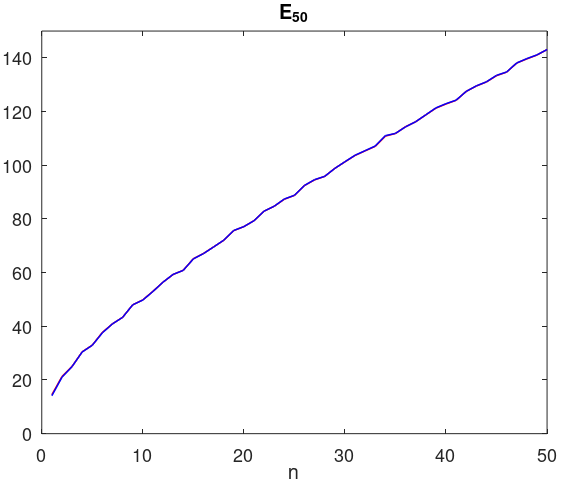}
\caption{The imaginary part of the first 50 nontrivial Riemann zeros and the eigenvalues calculated from the potential in FIG. \ref{potential50} .  Notice the 
very good agreement.}
\label{spectrum50}
\end{figure}

Finally, equation (\ref{v-reconstruct}) is used to calculate $V_{50}$, with the $C_n$-functions defined according to the above. This result can be compared to the direct numerical calculation
of $V_{50}$ in FIG. \ref{potential50}. The agreement is excellent! The same holds for the eigenvalue spectrum in FIG. \ref{spectrum50}.
I can also be observed that the fractal pattern in the potential, identified in \cite{wu-sprung} starts to appear.

Apart from the constant tail, the only parameter needed to define the correction function is the approximation error in the Riemann -- von Mangoldt formula. This is indeed quite remarkable!
It also leads to the intriguing possibility of formalizing the numerical experiments in this paper and hopefully gaining some new insights into the famous Riemann hypothesis itself.

\end{document}